# Voltage-induced strain clocking of nanomagnets with perpendicular magnetic anisotropies

Qianchang Wang, Jin-Zhao Hu, Cheng-Yen Liang, Abdon Sepulveda, Greg Carman*

*Department of Mechanical and Aerospace Engineering, University of California, Los Angeles, CA 90095, USA*

* Author to whom correspondence should be addressed: carman@seas.ucla.edu

**Abstract**

Nanomagnetic logic (NML) has attracted attention during the last two decades due to its promise of high energy efficiency combined with non-volatility. Data transmission in NML relies on Bennett clocking through dipole interaction between neighboring nanomagnetic bits. This paper uses a fully coupled finite element model to simulate Bennett clocking based on strain-mediated multiferroic system for Ni, CoFeB and Terfenol-D with perpendicular magnetic anisotropies. Simulation results demonstrate that Terfenol-D system has the highest energy efficiency, which is 2 orders of magnitude more efficient than Ni and CoFeB. However, the high efficiency is associated with switching incoherency due to its large magnetostriction coefficient. It is also suggested that the CoFeB clocking system is slower and has lower bit-density than in Ni or Terfenol-D systems due to its large dipole coupling. Moreover, we demonstrate that the precessional perpendicular switching and the Bennett clocking can be achieved using the same strain-mediated multiferroic architecture with different voltage pulsing. This study opens new possibilities to an all-spin in-memory computing system.

Currently computer information processing is based on CMOS (complementary metal-oxide-semiconductor) transistors. Nanomagnetic logic (NML) has emerged as a potential replacement of CMOS transistor with ultra-low energy dissipation.[1–3] In NML, bi-stable magnetic states are encoded as '0' and '1', which are non-volatile and theoretically require zero standby energy. The information is transferred using unidirectional magnetization propagation with dipole interaction between neighboring bits referred to as Bennett clocking.[3–5] Researchers have also demonstrated logic gate Bennett clocking designs,[1,6,7] however, generating a sufficient clocking field remains the primary challenge.

The common way to generate clocking fields is with an oscillating magnetic field either from external application[8,9] or on-chip generation by current through a wire[10]. Researchers have also experimentally demonstrated Bennett clocking using spin Hall effect.[11] However, both methods require high energy input (~100 fJ per flip[3]), thus sacrificing the intended advantage of NML, i.e. low energy dissipation. An alternative approach uses a strain-mediated multiferroic system representing an energy efficient technique to control nano-scale magnetic anisotropies.[12–15] Strain-mediated Bennett clocking has been demonstrated by both simulation[3,6,7,16–18] and experiment[19]. While the energy efficiency (~100 aJ per flip[3,19]) of strain-mediated Bennett clocking has been demonstrated, these studies are limited to in-plane magnetic system and analysis uses an oversimplified uncoupled macrospin model to understand the process. Compared to in-plane bits, nanomagnets with perpendicular magnetic anisotropy (PMA) are more promising due to the smaller bit size producing higher information density.[20,21] Furthermore, the dipolar interaction between PMA dots is less susceptible to shape variation, which is suggested to significantly impact device behaviors for in-plane Bennett clocking[8,22–24]. Therefore, theoretical examinations of perpendicular Bennett clocking system are needed to assess this concept and guide future NML design.

In this paper, we study a four-bit Bennett clocking system for different materials (Ni, CoFeB, and Terfenol-D) with perpendicular magnetic anisotropies. A 3D finite element model that couples micromagnetics, electrostatics, and elastodynamics are used to simulate the strain mediated Bennet clocking. The model assumes linear elasticity, linear piezoelectricity, and electrostatics. Thermal fluctuations are not considered in this model. Previous research has shown that the presence of thermal fluctuation at room temperature will increase switching error rate, which could be compensated by increasing the applied strain level.[7,25] The magnetic dynamics are governed by the Landau-Lifshitz-Gilbert (LLG) equation:[26]

$$\frac{\partial \boldsymbol{m}}{\partial t} = -\gamma(\boldsymbol{m} \times \boldsymbol{H}_{eff}) + \alpha \left(\boldsymbol{m} \times \frac{\partial \boldsymbol{m}}{\partial t}\right) \qquad (1)$$

where $\boldsymbol{m}$ is the normalized magnetization, $\mu_0$ is the vacuum permittivity, $\gamma$ is the gyromagnetic ratio and $\alpha$ is the Gilbert damping parameter. $\boldsymbol{H}_{eff}$ is the effective magnetic field defined by $\boldsymbol{H}_{eff} = \boldsymbol{H}_{ex} + \boldsymbol{H}_{Demag} + \boldsymbol{H}_{PMA} + \boldsymbol{H}_{ME}$, where $\boldsymbol{H}_{ex}$ is the exchange field, $\boldsymbol{H}_{Demag}$ the demagnetization field, $\boldsymbol{H}_{PMA}$ the effective PMA field, and $\boldsymbol{H}_{ME}$ the magnetoelastic field. The magnetocrystalline anisotropy is neglected by assuming the magnetoelastic material's grain size is smaller than its exchange length. The PMA field is expressed using a generalized equation:

$$\boldsymbol{H}_{PMA} = -\frac{2}{\mu_0 M_S} K_{PMA} m_z \hat{\boldsymbol{z}} \qquad (2)$$

where $K_{PMA}$ is the PMA coefficient.[15,27]

The magnetoelastic field $\boldsymbol{H}_{ME}$ is represented by the following equation[26]:

$$\boldsymbol{H}_{ME}(\boldsymbol{m}, \boldsymbol{\varepsilon}) = -\frac{1}{\mu_0 M_S}\frac{\partial}{\partial \boldsymbol{m}}\{B_1[\varepsilon_{xx}\left(m_x^2 - \frac{1}{3}\right) + \varepsilon_{yy}\left(m_y^2 - \frac{1}{3}\right)$$
$$+\varepsilon_{zz}\left(m_z^2 - \frac{1}{3}\right)] + 2B_2(\varepsilon_{xy} m_x m_y + \varepsilon_{yz} m_y m_z + \varepsilon_{zx} m_z m_x)\} \qquad (3)$$

where $m_x$, $m_y$ and $m_z$ are components of normalized magnetization along $x$, $y$ and $z$ axis, $B_1$ and $B_2$ are first and second order magnetoelastic coupling coefficients. $B_1$ and $B_2$ are calculated using the equation: $B_1 = B_2 = \frac{3E\lambda_S}{2(1+v)}$, where E is the Young's modulus and $\lambda_S$ is the saturation magnetostriction coefficient of the magnetic material. In the formula of $\boldsymbol{H}_{ME}$, $\boldsymbol{\varepsilon}$ is the total strain that consists two parts: $\boldsymbol{\varepsilon} = \boldsymbol{\varepsilon}_p + \boldsymbol{\varepsilon}^m$, where $\boldsymbol{\varepsilon}_p$ is the piezostrain calculated from linear piezoelectric constituitive equation, and $\varepsilon_{ij}^m = 1.5\,\lambda_s(m_i m_j - \delta_{ij}/3)$ is the strain contribution due to isotropic magnetostriction, $\delta_{ij}$ is Kronecker function[26]. The magnetization affects strain through $\boldsymbol{\varepsilon}^m$, and the strain affects the magnetization through $\boldsymbol{H}_{ME}$ term in equation (1). In other words, this analysis includes the full coupling between the mechanics and magnetics (or bidirectional), which is shown to be more accurate than one-way coupled simulation.[28,29] More details about the weak form development, the equations used, and the solvers implemented can be found in Liang et al.[28,30]

Figure 1(a) illustrates the simulated multiferroic structure consisting of a piezoelectric thin film on a substrate, magnetoelastic disks, and ground/surface electrodes. The piezoelectric material is PZT-5H[31] poled along the $z$ direction with 1000 nm × 1000 nm lateral $x$-$y$ dimension and a 100 nm thickness. The PZT film's four sides and bottom surfaces are mechanically fixed while the bottom surface is electrically grounded.

In this study, three different magnetoelastic materials with perpendicular magnetic anisotropy are investigated, i.e. Ni, CoFeB and Terfenol-D. For each material system, an array of four disks along the $x$ axis is simulated. All magnetoelastic disks have a 50 nm diameter, and their bottom surfaces are perfectly adhered to the PZT thin film. The thicknesses of the magnetoelastic disks depend upon the material modeled, as shown in Table I. The thickness values are chosen to ensure the magnetic state is thermally stable with a thermal energy barrier $\Delta E_b > 40 k_B T \approx 0.2 aJ$, for each material system studied. Each magnetoelastic disk is surrounded (along $y$ axis) by a pair of square electrodes with 30 nm side lengths. For each electrode pair, voltage is always applied or removed simultaneously while the bottom electrode is held grounded. The edge-to-edge distance from each magnetic disk and its control electrode is 20 nm. The edge-to-edge distance between neighboring magnetic disks (i.e., $d_{E-E}$ in Fig. 1(a)) depends on the material system, as shown in Table I. The $d_{E-E}$ is selected so that the dipole coupling between neighboring disks is sufficient for clocking while the magnetic interactions from other disks is negligible. The material parameters for Ni[26–28,32], CoFeB[21,33–35] and Terfenol-D[36–38] are provided in Table I. The Gilbert damping α for all materials are set to be 0.5 to improve numerical stability. The actual damping factors are: α(Ni) = 0.038[26], α(CoFeB) = 0.01[21], α(Terfenol-D) = 0.06[36], which are smaller than the damping used in the simulation. Therefore, the actual clocking speed is expected to be slower as it will take longer to stabilize at the transient in-plane state when the strain is turned on.

Fig 1(b) is the schematic of information flow for a four-bit nanomagnetic logic (NML) system. The information is encoded as the perpendicular magnetization $m_z$, which is illustrated by the arrow attached to each disk. Assume the four memory bits start as an anti-parallel magnetic state "↑↓↑↓" as shown in the first row in Fig 1(b). Initially new information is written in disk 1, and its magnetization is switched from up to down using a short (~< 1 ns) voltage pulse[15,27], as shown in the second row in Fig 1(b). When disk 1 changes its state, disk 2 does not spontaneously update its state because dipole coupling is insufficient to overcome the energy barrier of disk 2. Therefore, an additional clocking field is needed which is achieved by applying the same voltage to disk 2 and disk 3 (see the third row in Fig 1(b)). The voltage-induced strain produces a localized clocking field that rotates the disks' easy axes to in-plane. Then removing the voltage from disk 2 produces an unstable in-plane magnetic state susceptible to external dipole fields. However, disk 3 is still mechanically strained and its in-plane magnetization has a much smaller impact on disk 2 compared to disk 1. Therefore, the magnetization of disk 2 preferably aligns anti-parallel to the disk 1, which is "↑" as shown in the fourth row in Fig. 1(b). In other words, the magnetic state or information in disk 1 is now propagated to disk 2. The process is subsequently executed on the remaining magnetic bits (see the last two rows in Fig. 1(b)). This causes information from the input bit to cascade along the information line uni-directionally with the end-system exhibiting the opposite state "↓↑↓↑" to the initial state.

Figure 2 shows simulation results for the Bennett clocking process in a four Ni disk system with an edge-to-edge distance between neighboring disks $d_{E-E}$ of 50 nm and an initial perpendicular magnetic state represented as "↑↓↑↓". Fig. 2(a) plots the normalized average perpendicular magnetization $m_z$ (solid line) as well as the applied voltage (orange dashed line) as a function of time for each disk. A 3V (30 MV/m) is applied to disk 1 with 0.8 ns duration, which includes 0.1 ns ramping time for both voltage application/removal steps. The voltage-induced strain is tensile along $x$ axis and compressive along $y$ axis, producing an effective field $H_{ME}$ along the $y$ axis due to the negative $\lambda_S$ for Ni. The magnetization starts to rotate towards in-plane, and the voltage is removed when the magnetization crosses the $x$-$y$ plane (corresponding to $m_z = 0$). Then magnetization continues to precess to its new perpendicular equilibrium state $m_z = -1$, i.e. disk 1 undergoes 180° perpendicular switching. After disk 1 has stabilized at t = 2 ns, a voltage pulse (3V) is consecutively applied to disks 2, 3, and 4, for t = 2~4 ns, 3.5~5.5 ns, 5~7 ns, respectively. These 2 ns clocking voltage pulses are sufficiently long to stabilize the magnetization in-plane. It is worth noting that in real Bennett clocking system, longer pulses are needed for magnetization to stabilize in-plane, because the actual Gilbert damping is smaller than that is used in the simulation. Upon removal of the clocking voltage (t= 4, 5.5, 7 ns), each disk (2,3, and 4) flips to a new state that is anti-parallel to the preceding disks orientation due to the dipolar field as illustrated in Fig. 1(b). At t = 9 ns, the 4-disk system reaches a new equilibrium state with each disk having an opposite magnetic state to its initial state, and strain-mediated Bennet clocking is finished.

Figure 2(b) provides the magnetic spin configurations for disk 2 at four distinct times (t = 1, 3, 5, 8 ns) during its 180° switching. The red arrows represent the direction and amplitude of the in-plane magnetization components while the color contour quantifies the $m_z$ component's magnitude. The switching process is predominantly coherent, as shown by the uniformity of both contour color and arrow directions. The switching coherency is quantitively examined in Fig. 2(c) by plotting the temporal evolution of averaged magnetization amplitude for disk 2, which is defined as:

$$|<m>| = \sqrt{<m_x>^2 + <m_y>^2 + <m_z>^2} \qquad (3)$$

where $<m_x>$, $<m_y>$, $<m_z>$ denote the volume averaged magnetization in x, y, z directions, respectively. The $|<m>| = 1$ corresponds to complete coherent switching, where all the spins point in the same direction throughout the switching process. $|<m>| = 0$ represents a completely random spin switching process, where $<m_x>$, $<m_y>$, $<m_z>$ magnitudes are all zeros. As shown in Fig. 2(c), the Bennett clocking process for disk 2 (as well as the other disks) is very coherent during the whole Bennett clocking process.

Figure 3 shows Bennet clocking results for a CoFeB system with a thickness of 1.6 nm and $d_{E-E}$ of 70 nm. The larger $d_{E-E}$ relative to Ni is related to the substantially larger CoFeB Ms producing larger dipolar fields. Additionally, since CoFeB has a positive $\lambda_S$, the applied voltage produces an effective **$H_{ME}$** along x axis. As shown in Fig. 3(a), the 3.5 V used is similar to Ni because their magnetostriction coefficients are of similar magnitude. For CoFeB, the initial voltage pulse duration applied to disk 1 is 0.7 ns to produce 180° precessional switching. This is followed by consecutive voltage pulse widths of 2.5 ns duration applied to disks 2, 3, and 4 at t = 2~4.5 ns, 3.5~6 ns, 5~7.5 ns, respectively. The pulse duration is longer than Ni and the reason is explained as follows. At t = 4 ns, disk 2 and 3 have experienced 2 ns and 0.5 ns voltage/strain, respectively. Disk 2's magnetization has stabilized in-plane and is ready for voltage removal, however, disk 3 still has small precession near its temporal equilibrium state. If the voltage applied to disk 2 is removed at t = 4 ns, the small perturbation of disk 3 may cause a flipping error in disk 2. To avoid this, the voltage to disk 2 is applied until 4.5 ns when disk 3 is completely stabilized. This issue is not present in Ni system due to weaker dipole coupling making it less susceptible to small spin perturbations compared to CoFeB. In addition, taking into consideration the actual Gilbert damping of CoFeB (0.01) is smaller than that of Ni (0.038), the Bennett clocking process for CoFeB is expected to be slower than that for Ni system.

Figure 3(b) provides representative spin configurations for disk 2 at t = 1, 3, 5, 8 ns in the Bennett clocking process for CoFeB system. Similar to Ni system, the voltage-induced strain is tensile along the *x* axis and compressive along the *y* axis. However, the effective field $H_{ME}$ is now along the *x* axis due to the positive $\lambda_S$ for CoFeB. Therefore, the spins are aligned along *x* axis at t = 3 ns and 5 ns. The color non-uniformity present at t = 5 ns indicates the switching process possesses some incoherency relative to Ni. As shown by |<m>| for disk 2 in Fig. 3(c), slight incoherency is observed in CoFeB system, which is attributable to the relatively smaller exchange length of CoFeB compared to Ni (see Table 1). The incoherency is initially observed during voltage application and becomes larger upon removal of the voltage. This can be explained as follows. The dominating effective fields in the beginning and the end of the clocking, when $m_z$ is large, are both PMA field since $H_{PMA} \propto m_z$. $H_{PMA}$ is uniform throughout the disk, so the switching is highly coherent. During voltage application, the dominating field becomes $H_{ME}$, but $H_{ME}$ has a spatial distribution caused by a non-uniform strain generated from the patterned electrodes. This non-uniform strain leads to a spatial distribution of spins inside the disk. After removing the voltage (i.e. strain) applied to disk 2, the dominating effective field becomes the dipolar field but there is still the presence of an $H_{ME}$ due to strain generated from disk 3. Both $H_{ME}$ and dipolar field are spatially variant and thus, they both contribute to the magnetic incoherency. Therefore, the incoherency becomes even larger after voltage removal due to this combined effect, i.e. non-uniform dipolar fields and $H_{ME}$.

Figure 4 shows the Bennett clocking results for Terfenol-D system with a thickness of 2 nm and a $d_{E-E}$ distance of 60 nm. As shown in Fig. 4(a), the applied voltage for this system of disks is only is -0.3 V, which is an order of magnitude smaller compared to either Ni or CoFeB. This is directly attributed to the fact that Terfenol-D has the largest $\lambda_S$ amongst these three materials and represents the largest value available at room temperature in a soft magnetic material system. Initially a short -0.3 V pulse with duration of 0.4 ns is applied to disk 1 to achieve the precessional switching. This is followed by consecutive voltage pulses of 2 ns duration applied to the disks 2, 3, and 4 at t = 2~4 ns, 3.5~5.5 ns, 5~7 ns, respectively. This timing sequence is the same as Ni system and is attributed to similar magnitudes of dipolar coupling. Fig. 4(b) and 4(c) show the representative spin configurations and temporal evolution of |<m>| for disk 2 during the Bennett clocking process. The negative applied voltage induces compressive strain along *x* axis and tensile strain along *y* axis leading to an effective magnetoelastic field $H_{ME}$ along the *y* axis due to the positive $\lambda_S$ for Terfenol-D. It is important to note that the vertical temporal axis in Fig. 4(c) has a much larger range compared to the |<m>| plots for both Ni and CoFeB. Therefore, incoherency present in the Terfenol-D system is considerably larger than Ni and CoFeB. This is attributed to the much larger $\lambda_S$ thus stronger coupling to the non-uniform strain distribution as discussed in previous research.[15] Similar to CoFeB, the switching is incoherent when the voltage is applied disk 2, and becomes larger upon voltage removal.

In conclusion, strain-mediated Bennett clocking has been successfully performed in three popular magnetoelastic material systems. Ni has the most coherent clocking process. CoFeB encounters slightly incoherent switching, mainly due to its small exchange length. Terfenol-D exhibits larger incoherency due to large $\lambda_S$. This also suggests that $\lambda_S$ has a more important impact on magnetic coherency than exchange length. As a trade-off for incoherency, the Terfenol-D requires smaller voltage for clocking, producing 2 orders higher efficiency than either Ni and CoFeB systems, as shown in Table I. The energy dissipation per bit per flip is calculated as $E = \frac{1}{2}QV$, where $Q$ is the total charge on the pair of electric pads during voltage application, and $V$ is the applied voltage. While CoFeB is the most mature ferromagnetic metal in magnetic memory because of large readout signal in magnetic tunnel junctions, the large Ms of CoFeB requires increased distances between disks to avoid magnetic perturbation of adjacent disks. This sacrifices the memory density in CoFeB system. In contrast, Ni system has the potential to provide highest memory density. However, the on-chip readout mechanism for Ni or Terfenol-D is less mature than CoFeB. This study clearly reveals the strengths and shortcomings of different material systems in Bennett clocking for NML devices. Additional studies on hybrid NML combining different materials may be able to utilize advantages from each material system.

## Acknowledgements

This work was supported by NSF Nanosystems Engineering Research Center for Translational Applications of Nanoscale Multiferroic Systems (TANMS) Cooperative Agreement Award EEC-1160504.

## Author Contributions

Q.W. conceived and performed the simulations. C.Y.L. provided the macrospin modeling framework. G.C. directed the work. J.Z.H and A.S. discussed the results. Q.W. and G.C. wrote the manuscript. All authors commented the manuscript.

## Additional Information

**Competing Interests:** The authors declare no competing interests.

TABLE I. Parameters for Ni, CoFeB and Terfenol-D used in the model.

| Parameter | Description | Units | Ni | CoFeB | Terfenol-D |
|---|---|---|---|---|---|
| $t$ | Thickness | nm | 2 | 1.6 | 2 |
| $M_s$ | Saturation magnetization | A/m | $4.8 \times 10^5$ | $1.2 \times 10^6$ | $8 \times 10^5$ |
| $A_{ex}$ | Exchange stiffness | J/m | $1.05 \times 10^{-11}$ | $2 \times 10^{-11}$ | $9 \times 10^{-12}$ |
| $L_{ex}$ | Exchange length | nm | 8.52 | 4.70 | 4.73 |
| $\lambda_s$ | Saturation magnetostriction coefficient | ppm | -34 | 50 | 1200 |
| $E$ | Young's modulus | GPa | 180 | 160 | 80 |
| $\rho$ | Density | kg/m$^3$ | 8900 | 7700 | 9210 |
| $K_{PMA}$ | PMA coefficient | J/m$^3$ | $-1.3 \times 10^5$ | $-8.1 \times 10^5$ | $-3.4 \times 10^5$ |
| $V$ | Applied voltage | V | 3 | 3.5 | -0.3 |
| $\Delta t_p$ | Pulse duration | ns | 2 | 2.5 | 2 |
| $d_{E-E}$ | Edge-to-edge distance between neighboring disks | nm | 50 | 70 | 60 |
| $E_{flip}$ | Energy per flip | fJ | 11.1 | 14.5 | 0.11 |

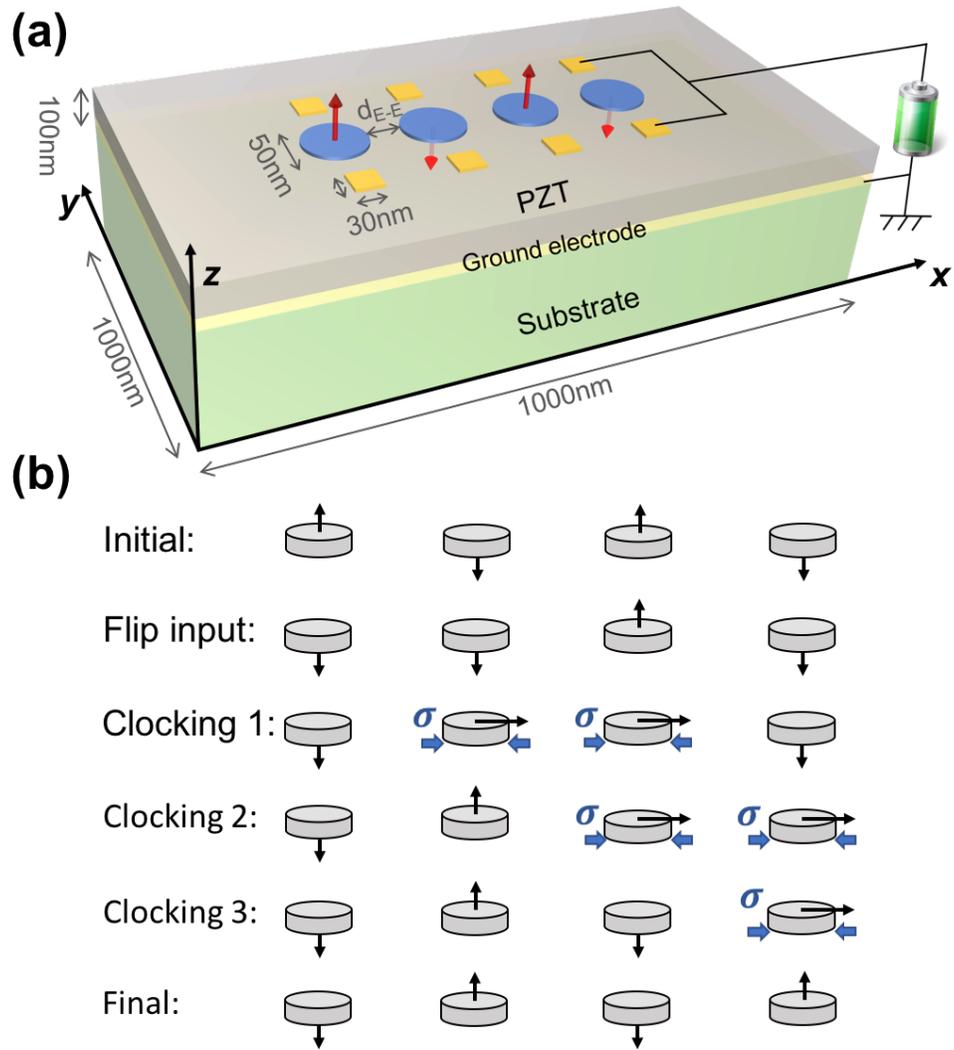

Fig. 1. (a) 3D illustration of the Bennett clocking system simulated in the model. (b) Information flow of Bennett clocking process.

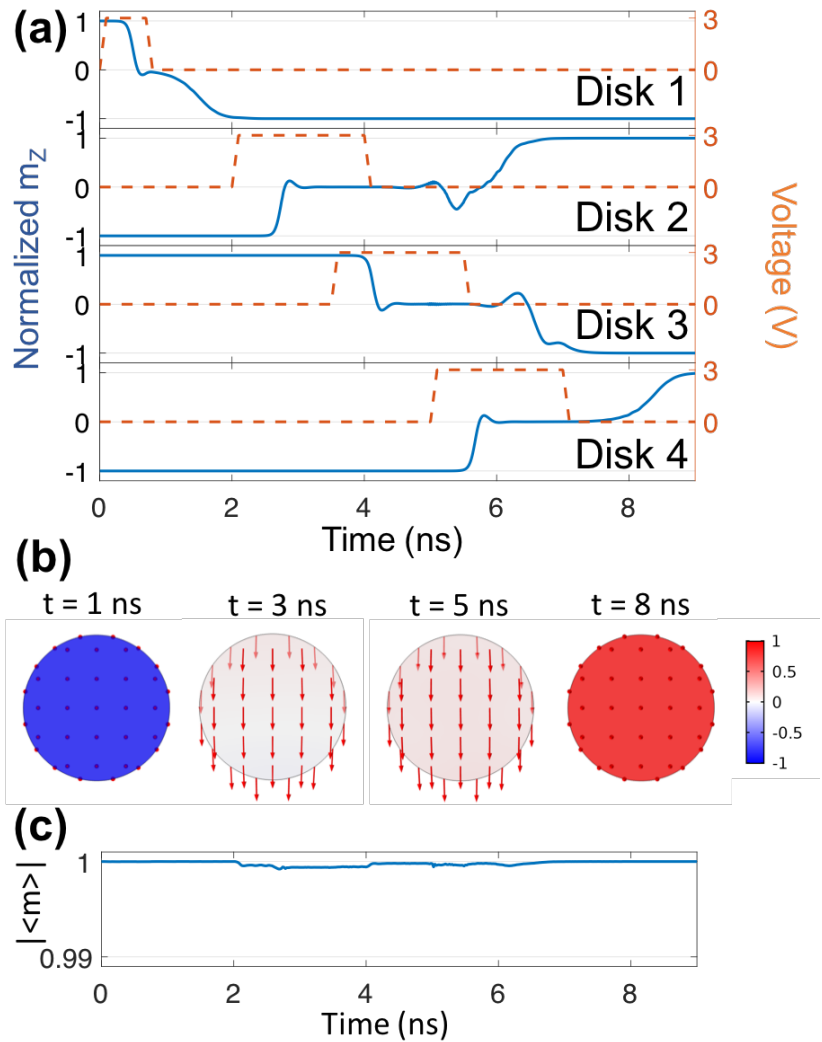

Fig. 2. Simulation results for Bennett clocking of a four-bit Ni system. (a) Temporal evolution of perpendicular magnetization and voltage application for the four Ni disks. (b) Vector diagrams of the magnetization distribution at t = 1, 3, 5, and 8 ns. The arrows represent the in-plane magnetization amplitude and direction, while the colors represent the perpendicular magnetization mz. (c) Coherency plot for disk 2.

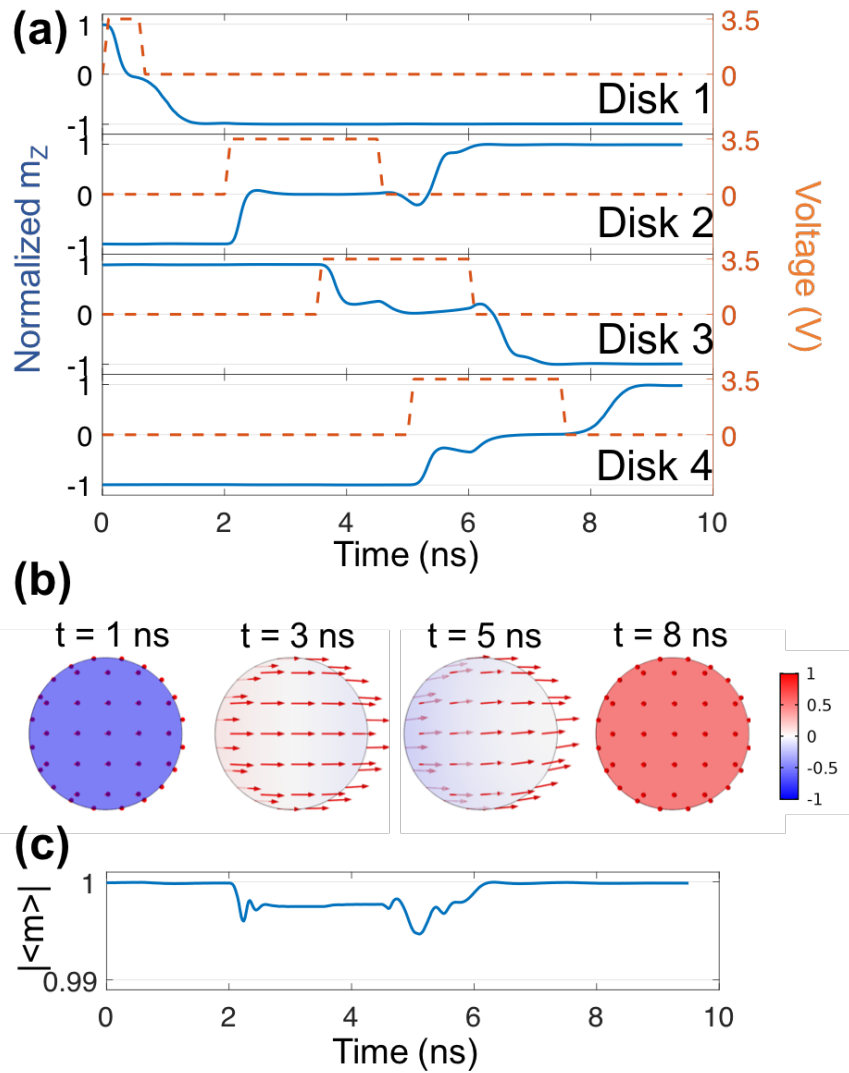

Fig. 3. Simulation results for Bennett clocking of a four-bit CoFeB system. (a) Temporal evolution of perpendicular magnetization and voltage application for the four CoFeB disks. (b) Vector diagrams of the magnetization distribution at t = 1, 3, 5, and 8 ns. (c) Coherency plot for disk 2.

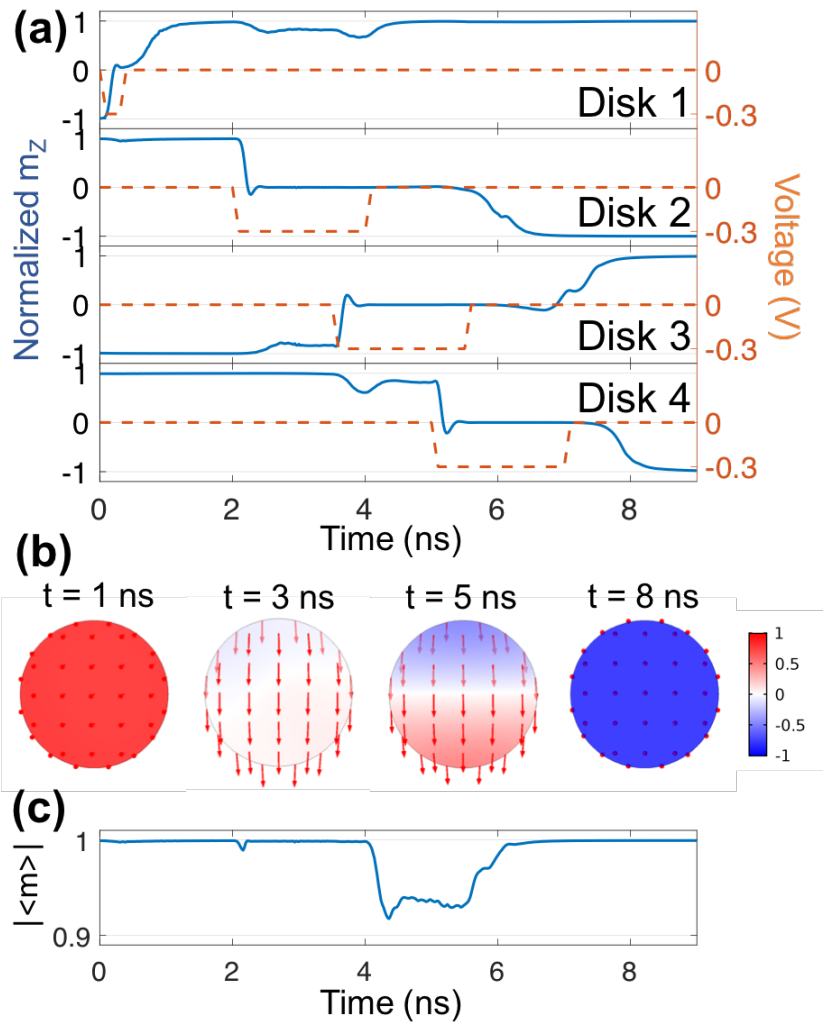

Fig. 4. Simulation results for Bennett clocking of a four-bit Terfenol-D system. (a) Temporal evolution of perpendicular magnetization and voltage application for the four Terfenol-D disks. (b) Vector diagrams of the magnetization distribution at t = 1, 3, 5, and 8 ns. (c) Coherency plot for disk 2.